\documentclass[12pt]{article}
\usepackage{a4wide,epsfig}
\begin{document}

\begin{center}

\vskip \baselineskip
{\Large\bf From the Skyrme model\\ to hypothetical Skyrmion stars:\\ 
 Astrophysical implications} \\
\vskip 3\baselineskip

R.\ Ouyed\,\footnote{Email: ouyed@phas.ucalgary.ca}
\vskip \baselineskip
{\small {\it Department of
Physics and Astronomy, University of Calgary, 2500 University Drive NW, Calgary, Alberta, T2N 1N4 Canada}}\\

\vskip \baselineskip

\end{center}

\vskip \baselineskip

\begin{abstract}

We discuss the Skyrme model for strong interactions
and the concept of a Skyrmion fluid. The 
 corresponding compact objects, namely Skyrmion stars,
constructed from the equation of state describing such a fluid
are presented and 
compared to models of neutron stars
based on modern equation of states.
We suggest plausible Skyrmion
star candidates in the {\bf 4U 1636$-$53} and {\bf 4U 1820$-$30}
 low mass X-ray binary systems
where the suggested masses of the accreting
compact companion ($\sim 2.0M_{\odot}$) remain
a challenge for neutron star models.

\end{abstract}

\section{Introduction}
\label{one}

We are closing in on neutron stars both observationally and theoretically.
Observationally more and better masses  and radii are
determined by a number of different methods.
The ``large" masses 
implied in few cases (as in the Vela X-1 pulsar
where a mass of $\sim 1.78 M_{\odot}$ has been suggested;
\cite{barziv01})
poses a real challenge to models of neutron stars
built using the so-called modern equation of state
(EOS) where the uncertainties are reduced by improved 
two- and three-body forces, relativistic effects
and many-body calculations \cite{heiselbergb00}. 
Even the stiffest EOS so far developed seem to be facing 
difficulty in accounting for the extreme values
(up to $2.4M_{\odot}$; Sect. 4).  
In \cite{ouyed99}, hereafter OB,
as an alternative, we 
constructed an EOS of
dense matter based on the Skyrme model
for strong interactions which represents
baryons as solitons of classic pionic fields.
The resulting compact objects we named {\it Skyrmion stars} (SSs)
\cite{ouyed02} are intrinsically 
heavier (due to the stiffness
of the Skyrmion fluid; hereafter SF) than any other type of compact stars computed
using modern EOS.  
SSs can be  
as massive as $\sim 2.8M_{\odot}$ 
leading us to speculate (given the above
mentioned observations) that these might  
exist in nature\footnote{Skyrmion stars, like neutron stars,
are likely to be born with masses around $1.5 M_{\odot}$.
We expect only older SSs that have accreted enough mass
to reach these extreme masses.}. 
The paper is presented as follows: Sect. 2 is devoted
to the discussion of the Skyrme model and its link to 
Quantum-ChromoDynamics (QCD). The problem of the
{\it missing attraction} in the Skyrme model is 
described.
In Sect. 3, we 
discuss the role the dilaton (the glueball potential in QCD) could play in
curing such a problem by binding
Skyrmions together to form the SF. 
We end Sect. 3 by  reminding the reader of the basic
properties of the SF and the resulting
EOS used to construct models
of SSs.
The astrophysical
implications follow in 
Sect. 4 where we compare 
SSs to stars constructed using modern 
EOS of dense matter. We conclude in Sect. 5.

\section{The Skyrme model}

We first give a brief overview of the Skyrme model.
The fundamental principles are discussed
at a basic level.  The interested reader is
referred to Ref. \cite{bhaduri88} for a thorough 
introduction to the topic.

Skyrme constructed a
model of pion interactions consisting of a conventional model of weak meson
interactions plus an additional (higher-order) term thought to take into account
indirect effects of heavier mesons like the $\rho$-meson.
The now well-known Skyrme Lagrangian density is usually written as \cite{skyrme62a},
\begin{equation}
L_{\rm Skyrme} = L_{2} + L_{4}
\label{one}
\end{equation}
where,
\begin{equation}
L_{2} = {f_{\pi}^2\over 4} Tr(\partial_{\mu}U\partial^{\mu}U^{+})
\label{two}
\end{equation}
is the Skyrme term ($U$ is the chiral field and $f_{\pi}$ is interpreted as the pion decay
constant), and 
\begin{equation}
L_{4} = {1\over 32e^2} Tr([U^{+}\partial_{\mu}U,U^{+}\partial_{\nu}U]
[U^{+}\partial^{\mu}U,U^{+}\partial^{\nu}U])
\label{three}
\end{equation}
is the quadratic term introduced by Skyrme to keep the Skyrmion stable against
the Derrick instability \cite{skyrme62b} ($e$ is the Skyrme parameter). Skyrme found that
his model contained `topologically nontrivial' configurations 
(extended objects) of the meson fields, namely
topological solitons, which he identified as baryons. 
For twenty years
the Skyrme model was overshadowed by the tremendous success of QCD 
and only in the early 1980's after the establishment of its
link to low energy QCD that the model was revived.

\subsection{Skyrmions and QCD}

The success of QCD is limited to the high energy regime, while at low energy it remains virtually intractable. The reason
for this is that QCD has a running coupling constant $\alpha_{\rm s}$; it is a function
of momentum transfer, or distance. At short distances of the order of
0.1 fm or less (high energy and momentum transfer of several GeV) QCD is characterized by
a small enough $\alpha_{\rm s}$ that it is treated perturbatively. All of the
results obtained in this regime are consistent with experimental data. This is the
phase in which the relevant degrees of freedom are quarks and gluons and it is called
the asymptotic freedom phase. At large distances of the order of 1 fm or more (low energy
and momentum transfer of 1 GeV or less) $\alpha_{\rm s}$ is of the order
of unity and QCD is a nonperturbative theory. This is the confinement
phase in which  quarks are confined inside hadrons and the hadronic degrees of freedom
are more relevant. This phase, which is the most practical, is the
most mathematically complex. It should provide all properties of hadrons such as masses, sizes, 
magnetic moments, lifetimes, scattering properties and, in principle, all
nuclear phenomena.

The first major step to overcome this problem was taken by  $'$tHooft
\cite{thooft74}. He found that in the limit of a large
number of colors (large $N_{c}$), $1/N_{c}$ could be
used as an expansion parameter. In this limit,
QCD simplifies a great deal and $'$tHooft went on
to show that at large $N_{c}$, QCD is equivalent
to a local field theory of mesons and `glueballs' (bound
states of gluons, without quarks), with an effective interaction
between them of order $1/N_{c}$. The second  step was taken by Witten
\cite{witten83}.
Assuming confinement, he showed that  baryons in large
$N_{c}$ QCD behave much like solitons in
a weakly coupled local field theory of mesons. In this
limit, baryon masses scale as $N_{c}=1/g^2$, where
$g$ is the strength of the meson coupling, while baryon sizes
are of order 1. Solitons in weakly coupled theories
have masses that scale as $1/g^2$ and sizes that
tend to constants as $g$ tends to zero. Even though
the mesons are weakly interacting, the solitons
interact strongly as do baryons in QCD. 

The next natural step, it seems, is to derive the effective meson
Lagrangian from the fundamental QCD Lagrangian. This task, as it turned out,
is immensely difficult. Its achievement is equivalent to
the solution of the intractable original problem of
low energy nonperturbative QCD in the confinement phase.
Nevertheless, the form of the resulting effective Lagrangian is being
narrowed down under reasonable assumptions. Assuming chiral
symmetry is spontaneously broken in QCD (with the physical
pions as the resulting Goldstone bosons, and taking the low energy limit in which one expects
the Golstone bosons to dominate the dynamics) it has been shown that
the first term in the resulting low energy effective Lagrangian is 
\cite{karchev85}:
\begin{equation}
L_{\rm eff.} = {N_{\rm c}f_{\pi}^2\over 16} Tr(\partial_{\nu}U\partial^{\mu}U^{+})+...
\label{four}
\end{equation}
which is astonishingly similar to the Skyrme term (Eq.~\ref{two})! This picture, which emerged
from large-$N_{\rm c}$ QCD, is precisely what Skyrme had in mind long before
QCD. Further work showed that the similarities between the Skyrme model and to what is described as the
nature of mesons and baryons in large $N_c$ QCD is simply intriguing.

The Skyrme model, however, as it is build was known to 
predict an isospin independent spin-orbit force with the {\it wrong} sign.
That is, it predicts a {\it repulsive} interaction.

\subsection{Skyrme model and the missing attraction}

The product ansatz for the two-baryon system as suggested in 
Ref. \cite{skyrme62a} (Eq.~\ref{two}), beyond its relative
simplicity as compared to other two-baryon field configurations which can be found in the
literature \cite{nyman86,amado93}, becomes exact for large $N-N$ separation. 
 Unfortunately, it is not the case for the isoscalar component of the spin-orbit force since the standard Skyrme
model predicts an isospin independent spin-orbit force with the {\it wrong} sign.
Namely, it predicts a {\it repulsive} interaction while the phenomenological
Bonn potential \cite{machleidt87} as the Paris potential
 \cite{lacombe80} gives an {\it attractive} one.
 This came to be known as the problem of the {\it missing attraction}.
Extensions of the Skyrme model consisted on 
including higher-order terms in powers
of the derivatives of the pion field \cite{moussallam93,abada93}.
 Expressed in terms of an $SU(2)$ matrix $U$
which, as we have said characterizes the pion field, a six-order term corresponding to $\omega$-meson
exchange \cite{jackson85},
\begin{equation}
L_{6} = - {\beta_{\omega}\over 2\omega_{\omega}^2} B_{\mu}(U)B^{\mu}(U)
\label{five}
\end{equation}
where $B^{\mu} = \epsilon^{\mu\nu\alpha\beta}Tr\left( (\partial_{\nu}U)U^{+}
(\partial_{\alpha}U)U^{+}(\partial_{\beta}U)U^{+}\right) / 24\pi^{2}$ is the
baryon current, $m_{\omega}$ the $\omega$-meson mass and $\beta_{\omega}$ a dimensionless
parameter related to the $\omega\rightarrow \pi\gamma$ width, might
be a good candidate to solve the problem of the $N-N$ isoscalar spin-orbit force.
While it was believed that the inclusion of such a term leads to the correct
sign ({\it attractive} interaction) for the isoscalar spin-orbit potential
\cite{riska89,kalbermann95}, recent 
calculations (see Ref. \cite{abada96} for e.g.) proved that by taking into account the second part of of the
sixth-order term\footnote{In Ref. \cite{riska89} and
\cite{kalbermann95} the authors considered only one part of the interaction
due to the sixth-order term in their calculations and omitted
the second part which arises from the exchange current \cite{nyman87}.} the anomaly of the
Skyrme model ({\it repulsive} force instead of {\it attractive}) remains.
The treatment of the spin-orbit
part of the two-pion exchange potential within the Skyrme model 
needed to be improved  in order to correct the anomaly of that sign. 
Below we explain how and why the dilaton field was suggested
as a plausible cure.

\section{The Dilaton field in the Skyrme model}

\subsection{The Dilaton contribution and Skyrmion structure}

In Ref. \cite{kalbermann95} and \cite{kalbermann97} the authors explored the idea
of coupling the Skyrmion to the dilaton field.
This idea to {\it account for a scalar field} confines the Skyrmion
and provides the attractive term missing in the original Skyrme
formalism (see OB for more on this).

\subsection{The Skyrmion fluid and Skyrmion stars}

We  start by writing the energy of $N$ Skyrmions per unit volume (parameterized by
the density, $\rho_V$) at finite temperature. In 
the mean field approximation it is given by
(where we adopt natural units with $\hbar=c=1$),
\begin{eqnarray}
E_{\rm V} = 2 g_N\int{d^3p\over (2\pi)^3}E_{\rm p}(n_{\rm p}+\bar{n}_{\rm p})
+V_{\sigma}(\sigma_{0}) &&\nonumber\\
 -{1\over 2}e^{2\sigma_{0}}
m_{\omega_{0}}^2\omega_{0}^2 + g_{\rm V}\omega_{0}\rho_{\rm V}\ .
\label{six}
\end{eqnarray}
Here, $g_V$ is the strength of the coupling of the $\omega_{0}$-meson
(of mass $m_{\omega_{0}}$; $\omega_0=<\!\!\!\omega\!\!\!>$ is  the mean-field value) to Skyrmions  while
$g_N$ represents the isospin degrees of freedom ($g_N=1$ for neutron matter
 and $g_N=2$ for symmetric nuclear matter)\footnote{In the energy equation as given by Eq.~(\ref{six}), the $\rho$
meson coupling has been omitted which reduces the analysis to symmetric
nuclear matter only. As such in OB the isospin
degrees of freedom ($g_{\rm N}$) was taken as a free
parameter as to allow for the two regimes -- pure neutron matter and
symmetric matter -- to be explored.}. 
 $E_{\rm p}$ is the contribution to energy of a single Skyrmion  while 
$p$ is the Skyrmion's momentum.
 $n_{\rm p}$ and $\bar{n}_{\rm p}$ parameterize the single particle and anti-particle
distribution functions.
 The pressure of such an ensemble
 at $T=0$ is then simply given
by $P_{\rm V} = \rho_V^2\partial (E_V/\rho_V)/ \partial\rho_V$. One
finds that the contribution of the
vector meson field ($\omega_{0}$) to the pressure grows
with density ($\omega_0\propto\rho_{\rm V}$), and is positive;
the dilaton potential $V_{\sigma}(\sigma_{0})$ (where $\sigma_0=<\!\!\!\sigma\!\!\!>$ is  the mean-field value)
 gives a negative contribution
to the pressure, acting to bind the system into
a fluid; the SF. Once the EOS describing such a fluid
is derived (OB) the corresponding compact objects
(SSs) are then computed \cite{ouyed02}.

SSs in our picture are not boson/soliton stars
where the soliton is a global structure over the scale of the star but rather
form their constituent baryons as topological solitons
using pions fields. This is fundamentally different
from other ``Exotic" stars  which also follow from
solutions to an effective non-linear field theory of strong
forces (see also \cite{walhout88,walhout90}; \cite{heusler92} and
references therein). 
In what follows,
we conclude this letter by describing
astrophysical cases plausibly supporting
the existence of  SSs in nature.

\begin{figure}[t!]
\begin{center}
\psfig{file=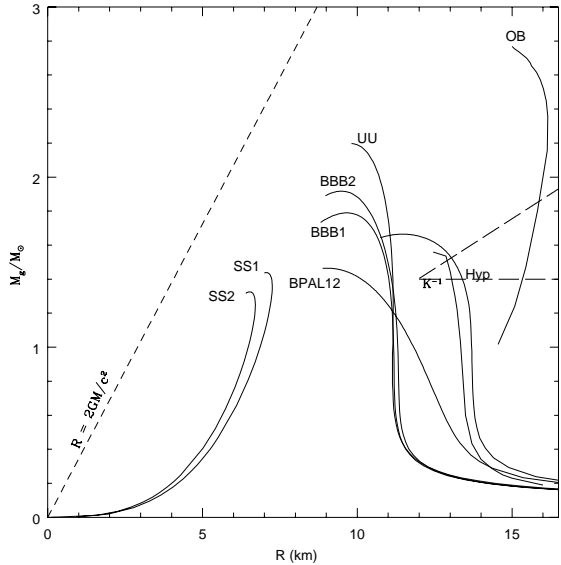,width=0.5\textwidth}
\caption{
The $M-R$ relation for non-rotating Skyrmion
stars (OB) as compared to  theoretical models of non-rotating neutron stars
(UU, BBB1, BBB2, BPAL12, Hyp, and K$^{-1}$) and
strange stars (SS1 and SS2);
see Ref. \cite{ouyed02}. 
The  Schwarzschild radius ($2GM/c^2$) is shown as a dotted line.
Inside the triangle is the allowed range of $M$ and $R$ for 4U 1636-53 as
modeled in Ref. \cite{nath02} using fits to X-ray bursts.}
\label{Fig.1}
\end{center}
\end{figure}

\section{Neutron stars vs Skyrmion stars}

\subsection{{\bf 4U 1636-53}}

In Fig.~\ref{Fig.1} we compare the $M-R$ relation for Skyrmion
stars (OB) to the  theoretical $M-R$ curve obtained using six recent
realistic models for the EOS (UU, BBB1, BBB2, BPAL12, Hyp, and K$^{-1}$).
The solid curves labeled SS1 and SS2 are for
strange stars (the data was kindly provided to us by the authors
of Ref. \cite{li99}).
The triangle depicts the mass-radius constraint from fits to X-ray bursts
in 4U 1636-53. Inside the triangle is the
allowed range of $M$ and $R$ which satisfies the compactness constraints
as modeled in Ref. \cite{nath02} (see their Figure 4), and
clearly favoring stiffer EOSs. Our modeled stars (OB) cross the triangle
suggestive of 4U 1636-53 as a plausible SS candidate.

\subsection{QPOs and Skyrmion stars: {\bf 4U 1820-30}}

{\it QPOs} are neutron stars emitting X-rays at frequencies of the
orbiting accreting matter. Such {\it quasi-periodic oscillations} (QPO)
have been found in 12 binaries of neutron stars with low mass companions.
If the QPO originate from the innermost stable
orbit \cite{zhang97,miller98}) of the accreting matter, their observed values
imply that the accreting neutron star has a mass of $\simeq 2.4M_{\odot}$
in the case of 4U 1820-30; this would
rule out most modern EOSs allowing only the stiffest ones. 

SSs is one possibility given that 
the gravitational mass
of the maximum stable non-rotating SS 
has a value of $\sim 2.8M_{\odot}$ (OB). 
For completeness,
one should note that even by making the modern/recent
EOS stiffer at high densities in a smooth
way, the maximum mass can never exceed $2.3M_{\odot}$ 
due to the causality condition 
\cite{heiselberga00}. 

\section{Conclusion}

We gave a brief historical overview of the
Skyrme model, its predictions
of hadronic interactions and its interesting connection
to QCD. Here we showed how the repulsive term in the
Skyrme model can be removed by coupling the  Skyrmion
to the dilaton field. This lead to the concept
of the Skyrmion fluid and the related hypothetical stars
we called Skyrmion stars. The stiffness of the Skyrmion
fluid allows for SSs to be as massive as $2.8M_{\odot}$.
The SSs show unique features; for a given mass their radii
are in general larger than those of neutron stars
constructed using modern EOSs.
 We discussed examples in astrophysics where SSs 
might constitute plausible candidates. Future observations
 constraining the mass-radius plane of compact stars would
most likely prove or rule out the existence of SSs in nature. 

\vskip\baselineskip
{\bf Acknowledgements.}
I am grateful to  
S. Morsink, G. K\"albermann and R. Bhaduri for encouraging help and valuable
discussions. The research of R.O. is supported by grants from the
Natural Science and Engineering Council of Canada (NSERC).

\end{document}